\documentclass[prl,twocolumn,preprintnumbers,amsmath,amssymb]{revtex4}
\usepackage{graphicx}% Include figure files
\usepackage{dcolumn}% Align table columns on decimal point
\usepackage{bm}% bold math
\usepackage{bibtopic}

\begin{document}

\def\Gbar{$\overline{\Gamma}$}
\def\Ef{$E_{\rm F}$}
\def\Ed{$E_{\rm D}$}
\def\Eg{$E_{\rm g}$}
\def\Efmath{E_{\rm F}}
\def\Edmath{E_{\rm D}}
\def\Egmath{E_{\rm g}}
\def\Tc{$T_{\rm C}$}
\def\kpara{{\bf k}$_\parallel$}
\def\kparamath{{{\bf k}_\parallel}}
\def\kperp{{\bf k}$_\perp$}
\def\invA{\AA$^{-1}$}
\def\Kbar{$\overline{\rm K}$}
\def\DeltaEf{$\Delta E_{\rm F}$}
\def\DeltaEfmath{\Delta E_{\rm F}}

%\preprint{APS}
  
\title{ Topological surface state under graphene for two-dimensional spintronics in air}

\author{
A. Varykhalov$^1$, D. Marchenko$^1$, M. R. Scholz$^1$, E. Rienks$^1$, T. K. Kim$^2$,
 O. Rader$^1$
}  
\affiliation{$^1$Helmholtz-Zentrum Berlin f\"ur Materialien und Energie, 
Elektronenspeicherring BESSY II, Albert-Einstein-Str. 15, D-12489 Berlin, Germany}
\affiliation{$^2$Institute for Solid State Research, IFW Dresden, P.O. Box 270116, 
D-01171 Dresden, Germany}

\begin{abstract}
{\bf
Spin currents which  allow for a dissipationless transport of information
can be generated by electric fields in semiconductor heterostructures 
in  the presence of a  Rashba-type 
spin-orbit coupling. The largest Rashba effects occur for electronic
surface states of metals but these  cannot exist but
under ultrahigh vacuum conditions.
Here, we reveal a giant Rashba effect ($\alpha_R \approx 1.5\cdot10^{-10}$ eVm)
on a surface state of Ir(111). We demonstrate that  its spin splitting and spin 
polarization remain unaffected when Ir is covered with graphene.
 The graphene protection is, in turn, sufficient for the spin-split surface state 
to survive in ambient atmosphere.
We discuss this result along with evidences for a  topological protection 
of the surface state. 
}
\end{abstract}  % 

%\pacs{}  

\maketitle

The Rashba effect in spintronics is based on symmetry breaking \cite{Rashba}:
In the bulk of crystallographically inversion symmetric solids  the 
time reversal symmetry results in degenerate spin subbands of electronic 
valence  states.
At crystal surfaces or  interfaces this structural inversion  symmetry
is broken and spin degeneracy is lifted. A gradient of the 
electric field perpendicular to the surface will then lead to a
Rashba effect 
in two-dimensional systems with large spin-orbit coupling \cite{Winkler}.
It emerges as a splitting of the band structure into  subbands 
$E_\pm(\kparamath)$ of opposite spin which are
displaced by  electron wave vectors $\pm\Delta$\kpara\   in opposite 
directions in momentum space. 
Regarding the transport properties of the solid and their applications,
 this leads to the generation of dissipationless spin currents without 
the necessity for ferromagnetic materials or magnetic fields 
\cite{MurakamiScience03,SinovaPRL04}.

In the quantum mechanical description of a two-dimensionally confined
electron gas, the Rashba effect is accounted for by the Hamiltonian
$$H_R= \alpha_R  \left( \sigma_x \frac{\partial}{\partial y} - 
                   \sigma_y \frac{\partial}{\partial x} \right) $$
where $\sigma$ denotes Pauli spin matrices and the parameter $\alpha_R$ 
is proportional to the potential gradient $\nabla V$ in $z$-direction and
accounts for the size of the spin-orbit interaction. 
For free electrons, the two spin-split bands $E_\pm(\kparamath)$ are described
by 
$$ 
E_\pm(\kparamath )=  \frac{\hbar^2 \kparamath^2}{2m^*} 
                        \pm \alpha_R \vert \kparamath \vert   ,
$$
 where $m^*$ is the effective mass. The two $E_\pm(\kparamath )$
parabolas are shifted
relative to the origin \Gbar\ ($\kparamath=0$) by a momentum splitting 
$\Delta \kparamath = ( m^* \alpha_R ) / \hbar^2 $ .

The Rashba effect has been investigated first and foremost for semiconductor
heterostructures \cite{Winkler}, reaching Rashba parameters of 
up to $4\cdot10^{-11}$ eVm in InAs-based structures \cite{Wu10}.
It has been shown that a Rashba-split system leads to an intrinsic
 spin Hall effect \cite{MurakamiScience03,SinovaPRL04}. 
An electric field applied along in the two-dimensional 
interface plane leads during charge transport to a tilting of the spins in opposite 
directions depending on the propagation direction 
\cite{MurakamiScience03,SinovaPRL04,online}. 
Spintronics does not require semiconducting materials because it does not depend
on charge currents and their amplification. 
Spin currents produced by the spin Hall effect in all-metal devices 
are much larger than in semiconductors \cite{TKimura,Seki}. 
Also, the largest Rashba effects have been measured for metal surface 
states which are localized at the outermost atomic layers of the solid-vacuum
interface. Discovered by Russian theoretician Igor Tamm in 1932 \cite{Tamm}, 
metal surface states  had to wait for their experimental scrutiny until
ultrahigh vacuum sample environments became available \cite{ReviewPlummer}.
They are highly sensitive to adsorbates, and it is hard to conceive an
adsorbate that will leave a surface state unaffected while protecing it 
against the influence of air.  

The most prominent example of the Rashba effect at metal surfaces is the L-gap 
surface state on Au(111) \cite{LaShell,Hedegard}.
% Henk
Spin- and angle-resolved photoelectron spectroscopy reveals a large 
spin-orbit splitting 
($\alpha_R = 3.3\cdot10^{-11}$ eVm) \cite{LaShell}.
Rashba-type splittings were also evidenced in  Bi \cite{bismuth}
and W(110) covered by monolayers of hydrogen \cite{Hochstrasser-PRL-2002}, Li
    \cite{RotenbergLiW110} Au, and Ag  \cite{Shikin-PRL-2008}. 
Large spin-orbit effects 
were recently observed 
in quantum well states of Au/W(110)
 ($\alpha_R \approx 1.6\cdot10^{-11}$ eVm)
\cite{Varykhalov-QWS-PRL-2008} and 
also  Pb/Si(111) 
($\alpha_R \approx 4\cdot10^{-12}$ eVm) \cite{Dil-PRL-2008}.
A  Rashba effect emerges also in the linear bands of massless Dirac 
fermions in  graphene on a Au monolayer   
  \cite{Varykhalov-Graphene-PRL-2008}.
The term {\it giant Rashba effect }
has been coined recently for a Rashba parameter $\alpha_R$ of the 
order of $10^{-10}$ eVm for a 
surface state of Bi/Ag(111) (
$\alpha_R \approx 4$--$6\cdot10^{-10}$ eVm) \cite{Ast}.
% 0.12 and 0.16 invA at RT and 77K
For all these states holds that they  cannot exist but in ultrahigh vacuum and, 
therefore, have no apparent practical application.

\begin{figure}[t]
\centering
\includegraphics[width=0.47\textwidth]{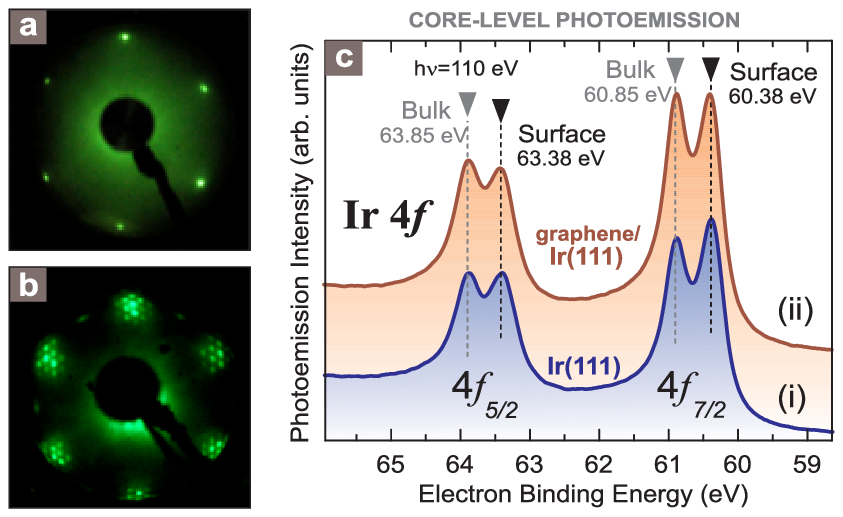}\\
\begin{flushleft}
{\bf Figure 1.} Preparation of epitaxial graphene.({\bf a})
LEED image of bare Ir(111) and ({\bf b}) of moir\'e-patterned 
graphene/Ir(111). ({\bf c}) Comparison of $4f$ core-level spectra measured 
from (i) bare Ir(111) and (ii) graphene/Ir(111). The surface component remains unchanged.
\end{flushleft}
\end{figure}

Figures 1 and 2 show the  characterization of clean and graphene-covered
Ir samples.  Bare Ir(111) shows a sharp $p(1\times1)$ pattern in 
low-energy electron diffraction (LEED). The LEED image of 
graphene/Ir(111) (Fig. 1b) reveals a characteristic diffraction pattern with 
multiple satellite spots which are due to the formation of a 
 superstructure of the moir\'e type which, in turn, is caused by a large   
 misfit ($\sim$10\%) between crystal lattices of graphene and Ir 
\cite{NDiaye-NJP-2008}. 
The high structural quality of the 
moir\'e-patterned graphene is also confirmed by direct microscopic 
characterization with scanning tunneling microscopy  \cite{online}.
  
Figure 1c shows the Ir $4f_{5/2}$ and $4f_{7/2}$ core levels, both with 
components originating from inside the bulk and from the surface.
The components  
at 0.5 eV lower binding energy are exclusively due to emission
from the topmost atomic layer of Ir(111) \cite{VeenSCLS}, and they were
shown by oxygen adsorption to be very sensitive to the local atomic environment
\cite{BianchiNJP09}.
It is, therefore,  remarkable that the spectra of the Ir$4f$ core level  
in Fig. 1c {\it cannot} be used as indicator 
for the presence of graphene on Ir(111). 
On the contrary,   the 
formation of graphene on top does not visibly affect the surface component
and line fits show that the energy splitting between bulk and surface components 
decreases   by only 2\%\  \cite{Lacovig-PRL-2009}.
%% 10 meV
This indicates very weak graphene-Ir interaction and is in line with
the observation that Ir(111) 
 is a metal substrate that supports the formation
of an ideal quasifreestanding electronic structure of graphene 
\cite{Pletikosic-PRL-2008}, similar to 
the substrate Au/Ni(111)   \cite{Varykhalov-Graphene-PRL-2008}.
 
Figs. 2a and 2b compare the overall valence band structure of Ir(111)
before and after formation of graphene, respectively.  For graphene/Ir(111) 
(Fig. 2b) one can see strongly dispersing graphene-derived $\pi$- 
and $\sigma$-states reaching at \Gbar\ binding energies of 8.38 eV and 3.65 eV, 
respectively.
(Note that the $\pi$- and $\sigma$-dispersions are accidentally degenerate 
with emission from Ir bulk bands.)
The high structural quality of the synthesized graphene is additionally
confirmed by the observation of a sharp conical dispersion of Dirac 
fermions at the $\overline{\rm K}$-point \cite{online}.

\begin{figure}[th]
\centering
\includegraphics[width=0.47\textwidth]{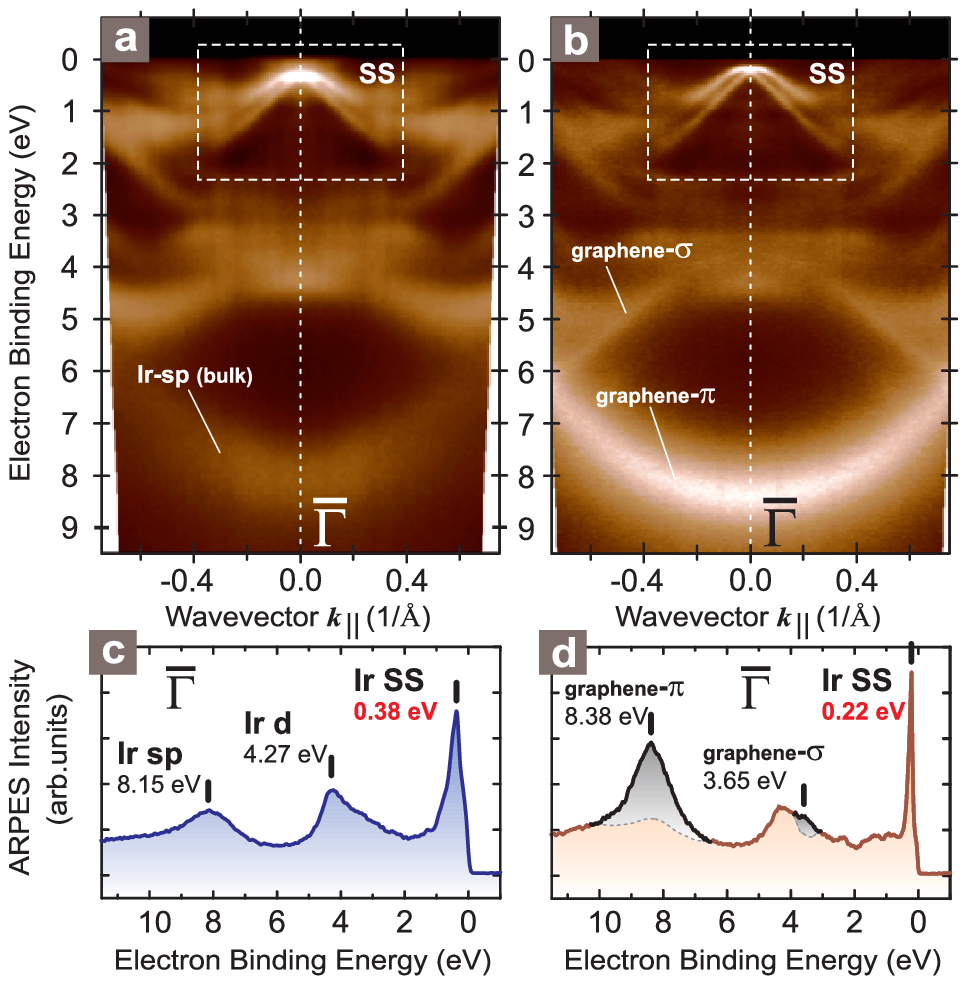}
\begin{flushleft}
{\bf Figure 2.} Overall valence band structure.   
({\bf a})   Bare Ir(111)   and  ({\bf b})  graphene-covered Ir(111)  (h$\nu=62$ eV).
Spin-orbit split surface state (SS) is marked with a white frame.
({\bf c, d})  normal-emission spectra [corresponding to the $\overline{\Gamma}$-point
of the surface Brillouin zone] extracted from ({\bf a}) and ({\bf b}), respectively. 
The spectra allow for a quantitative comparison between intensities of
Ir surface state and graphene bands.
\end{flushleft}
\end{figure}

\begin{figure*}[th]
\centering
\includegraphics[width=0.96\textwidth]{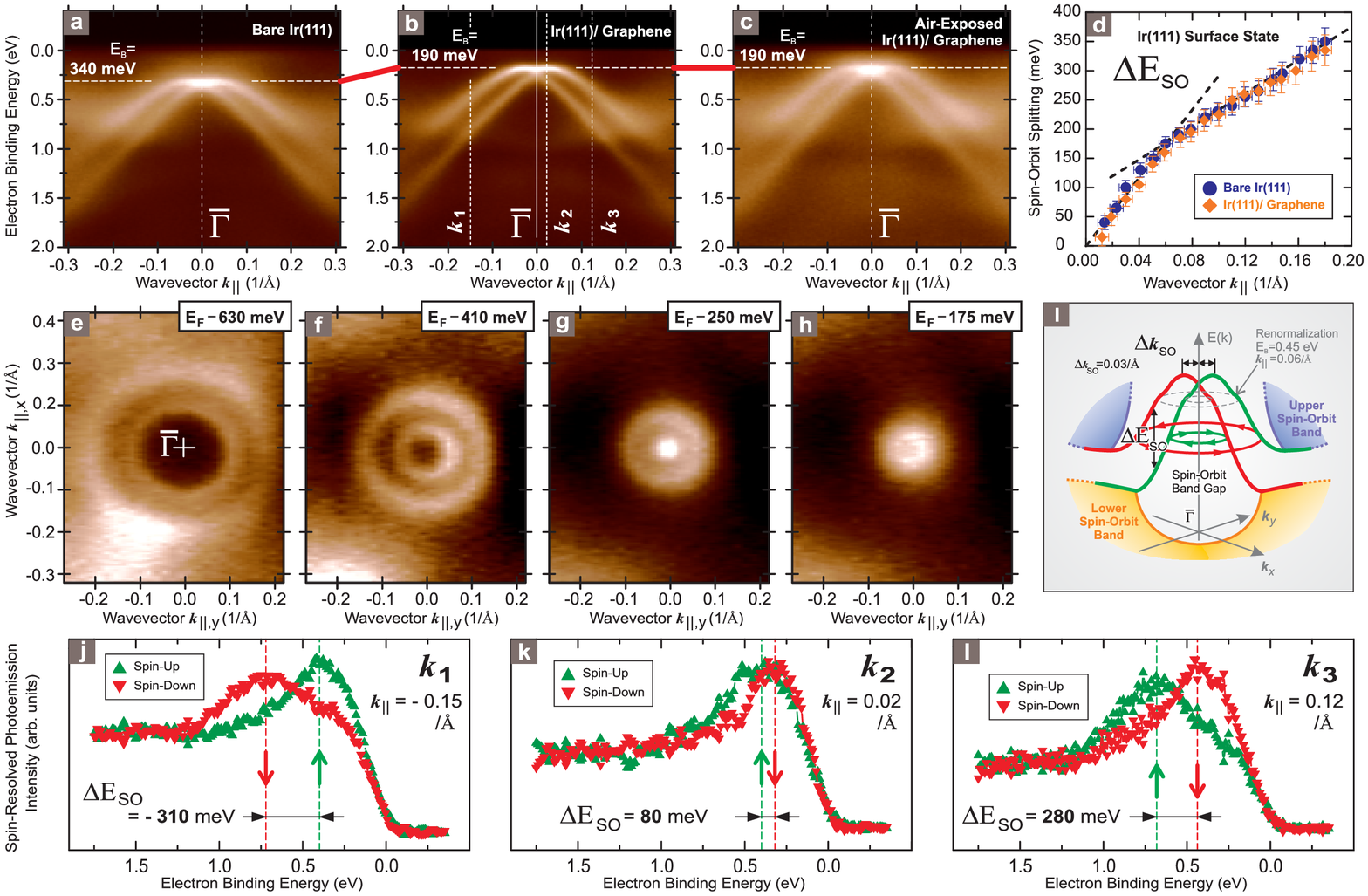}
%\caption{
\begin{flushleft}
{\bf Figure 3.}
Characterization of the spin-orbit split surface state on Ir(111). 
Dispersion of surface state on ({\bf a}) bare Ir(111) and ({\bf b})
under graphene grown on Ir(111).
({\bf c}) Dispersion of surface state under graphene on Ir(111) after 
exposure to ambient atmosphere for 15 min.
({\bf d}) Detailed comparison of spin-orbit splitting
$\Delta$E$_{\rm SO}$({\it k}$_\parallel$) for bare and graphene-covered Ir(111).
({\bf e--h}) Constant energy surfaces of  the Ir surface state extracted
from full-photoemission mapping around $\overline{\Gamma}$.
In agreement with the Rashba model,  two circles are revealed
corresponding to opposite directions of spin circulation.
({\bf i}) Sketch of spin-orbit splitting and spin-topology of the surface
state indicating topological protection. 
({\bf j--l}) Direct proof of spin-splitting in Ir surface state 
under graphene by spin-resolved photoemission. Measured spectra correspond to
wave-vectors {\it k}$_{1}$, {\it k}$_{2}$ and {\it k}$_{3}$ 
as denoted in ({\bf b}).
The photon energy was 62 eV.
\end{flushleft}
%}
\end{figure*}

The comparison between Figures 2a and 2b delivers another 
important message: Apparently, the Ir bands preserve their overall 
dispersions and binding energies upon graphene deposition. This 
agrees well with the weak interaction between graphene and Ir(111), 
noted above.
In this context, very intriguing is the band showing up 
at 0.2--0.3 eV below \Ef\ at \Gbar\  
[the region is marked by a frame and the dispersion is denoted as SS 
in Figs. 2a and 2b]. 
From studies at \Gbar\ this band is known to be an Ir(111) surface state 
 situated at 0.4 eV binding energy 
  above the bulk L$_{6-}$ point at 1 eV 
binding energy \cite{VeenSSPRB80}.  
The state is related to the L-gap surface states of Au(111)
but in Ir the bulk bands have a different order so that the surface state 
is reversed (i. e., with effective mass $m^*<0$). 
It is seen now that this surface state is also 
% from Ir bulk {\it d}-states at $\sim$0.7 eV and $\sim$1.2 eV
 clearly   split  in \kpara, resembling classical Rashba-type spin-orbit 
split surface states \cite{LaShell}.
The splitting amounts to $2\Delta\kparamath = 0.075$ \invA.
%  which is in line with the atomic number of $^{77}$Ir 
%  is similar to that of $^{79}$Au. 
The population of this state  remains unaffected 
by the formation of graphene on Ir(111). 
This can be seen quantitatively in normal-emission (\Gbar) spectra  
 plotted in Figs. 2c and 2d.
 We paid  attention explicitly  to the  splitting sampling the surface-state
region  with enhanced angular and energy resolution. The dispersion 
acquired for bare Ir(111) is shown in Fig. 3a. Indeed, the electronic
state is composed of two identical parabolic bands shifted by
$  \pm$0.04 \AA$^{-1}$ relative to 
$\overline{\Gamma}$. The binding energy of the top of the bands is 
determined as 340 meV.  Fig. 3b demonstrates what happens to
 the split-state when graphene is grown on top:
The only change concerns the binding energy of the parabolic 
bands, which decreases by 150 meV to 190 meV. 
% due to modification of surface potential by graphene.
(The different sharpness of the bands is not a systematic effect.)
This is clear evidence that the split band  is a {\it surface state}
of Ir(111) which persists {\it under} graphene.

Results of a  full photoemission mapping of the Ir 
surface state under graphene are presented in Figs. 3(e-h) 
as a sequence of constant 
energy surfaces cut at binding energies of 630, 410, 250 and 
175 meV. 
% conformative
The maps are fully consistent with the Rashba mechanism of  
spin-orbit splitting: For higher binding energies two circles
are observed which correspond to opposite directions of spin circulation
(arrowheads in Fig. 3i). 
At $\sim$250 meV, the inner circle shrinks to a point which is the Kramers
point of spin degeneracy, and for 175 meV binding energy
 only one circle is left, which
corresponds to the only direction of spin rotation.  
This mapping also shows that two bands  that   disperse in parallel
from the zone centre as have been obtained from Ir(111) recently
\cite{PletikosicJPCM10} can easily result from a slight misalignment 
of the experimental setup.

We have additionally confirmed
this Rashba scenario through direct observation of the spin.
Spin-resolved spectra measured for the emission angles corresponding to the electron
wave vectors {\it k}$_{1}$, {\it k}$_{2}$ and {\it k}$_{3}$ [labels of
Fig. 3c] are displayed in Figs. 3j--l. Indeed, the splitting of the  bands 
is a spin splitting. The splitting $\Delta E_{\rm SO}$ is large for $k_3>0$ 
far from \Gbar, decreases for $k_2$ closer to \Gbar, 
and, expectedly, reverses for $k_1<0$. 

We have fitted the surface-state dispersions  
shown in Figs. 3a and 3b and investigated how the spin-orbit splitting 
$\Delta E_{\rm SO}$ behaves with wave-vector {\it k}$_\parallel$
 for bare Ir(111) and for graphene-covered Ir(111). 
Fig. 3d shows that in both cases $\Delta E_{\rm SO} (k_\parallel)$
is linear, corresponding to parabolic, free-electron-like bands, 
and fulfills the criterion 
$\Delta E_{\rm SO} (k_\parallel=0)=0$ of the Rashba effect.
However, $\Delta E_{\rm SO}$ suffers a kink at 
$k_\parallel$=0.06 \AA$^{-1}$. This can be ascribed 
to the interaction of the surface state with Ir bulk $d$-bands in this
range because the surface states are partially degenerate with 
bulk bands \cite{VeenSSPRB80} and, are therefore rather surface resonances. 

The other surprising  information 
available from Fig. 3d is a {\it quantitative} equality of 
the Rashba splitting $\Delta E_{\rm SO}$ for bare Ir(111) and for 
graphene/Ir(111). 
According to a recent analysis  \cite{Giovannetti}
there is a large and universal charge gradient $\Delta V$ at  
any graphene-metal interface which can be estimated for graphene/Ir(111)
 from the work functions
of  Ir(111), 5.8 eV, and of graphene, 4.5 eV, and the Fermi-energy shift 
$\Delta$\Ef, 0.1 eV, as 1.2 eV.
The surface  potential  has been assigned an important role for the 
Rashba effect \cite{RotenbergLiW110,Hedegard,Hochstrasser-PRL-2002}.
Based on Li/W(110) adsorption experiments, 
the spin-orbit splitting has even been suggested as
local probe of surface potential gradients \cite{RotenbergLiW110}. 
Previously, we found no such influence for Au/W(110) and related
systems \cite{Varykhalov-QWS-PRL-2008}. 
Here, the change in surface potential modifies the binding energy but not
the spin-orbit splitting of the present surface state. 
 The spin-orbit effect is therefore assigned to the  large potential gradient of 
nuclear charge of the $^{77}$Ir atoms.
 
Before closing, we
 want to address the reason for the robustness of the 
surface state. Recently, topological surface states have
been predicted and observed which are spin-orbit split and
 protected by time-reversal symmetry \cite{topological,XiaNatPhys,HsiehScience09},
and it is the question whether the presently encountered stability 
of the surface state towards graphene deposition is related.
It was pointed out recently \cite{XiaNatPhys} that the 
spin-polarized and Rashba-split L-gap surface state of Au(111) does not
fulfil the criterion for topological protection, which is an odd number of 
Fermi level crossings between two time-reversal invariant k-points of the 
surface Brillouin zone, as occurs with the surface state identified on Bi$_2$Se$_3$  
\cite{XiaNatPhys}.
 Unlike on Au, the present Ir surface state becomes degenerate
with bulk $d$-bands,
and its tails  connect to bulk $d$   bands as seen in Fig 2. 
We can demonstrate experimentally that  
this state connects L$_{6-}$ and L$_{6+}$ bulk states of Ir.\cite{online}
While it apparently does not fulfil
the criterion for topological insulators 
of an odd number of crossings of \Ef\  
between the two time-reversal
invariant {\bf k}-points \Gbar\ and $\overline{\rm M}$, this criterion, 
if adopted to the metal Ir, has to consider
 instead of the Fermi level 
a curved line lying between these bulk bands, and such line is crossed
by the surface state in fact only an odd number of times 
between \Gbar\ and $\overline{\rm M}$.\cite{online}
A similar situation  has been found for the (111) surface of Sb recently.
\cite{HsiehScience09}

Finally, considering the ability of the spin-split surface state
to exist under graphene, we have tested how well  graphene  may
protect the Ir surface state from the environment. We have exposed a
graphene-covered Ir(111) sample to ambient atmosphere for 15 min
and measured the dispersion of the surface state right after this. 
The result is presented in Fig. 3c. Although a somewhat stronger background 
is seen due to remaining adsorbates, 
an accurate analysis shows that neither binding energy nor  
spin splitting of the surface state are influenced by the air (Fig. 3d).
In a technological context this means that graphene, weakly interacting 
with its substrate, can be considered  an ideal capping layer, 
which on the one hand protects metallic surfaces from
a chemically reactive environment \cite{PLDS}, and on the other hand
keeps its electronic and spin structure intact even at the atomic
level of surface electronic states which are most promising 
generators of two-dimensional spin currents. 

In summary, we have demonstrated  
(i) a previously unknown  Rashba-type splitting of giant size of a surface state 
on Ir(111), 
(ii)  that the   surface state survives when Ir(111) becomes
covered with epitaxial graphene while it changes its binding 
energy by 150 meV, 
(iii) that  neither dispersion nor Rashba splitting of the surface state
are influenced by the presence of graphene on top of Ir(111), 
(iv) that the spin-dependent band dispersion is consistent with topological
protection, 
(v) and that graphene protects the surface electronic structure of Ir(111) 
so well that the spin-orbit split surface state survives the exposure of the 
sample to ambient atmosphere.
% band gap engineering and reducing the film thickness

\phantom{xxx}

\noindent{\bf Methods}

\noindent For information on experimental methods and
 sample characterization please see
the Supplementary Information \cite{Supplement}.

\newpage

\onecolumngrid

\begin{center}
\large
\noindent
Supplementary information for\\
\vskip0.2cm
\noindent
{\bf Topological surface state under graphene for two-dimensional spintronics in air}\\
\normalsize
\vskip0.3cm
\noindent
A. Varykhalov, D. Marchenko, M. R. Scholz, E. Rienks, T. K. Kim, and O. Rader\\
\vskip0.5cm
\end{center}

\twocolumngrid

\begin{center}
{\bf Methods}
\vskip0.5cm
\end{center}

Experiments were conducted in ultrahigh vacuum better than 1$\times$10$^{-10}$ mbar. 
The clean Ir(111) surface was prepared by cycles of Ar$^{+}$ sputtering followed 
by annealing at 1600 K. The graphene layer was grown epitaxially by chemical 
vapour deposition of propene at 1150 K and a partial pressure of 3$\times$10$^{-8}$ mbar. 
Spin- and angle-resolved photoemission measurements were performed with 
electron analysers Scienta R8000 (setup ARPES 1$^{2}$) and SPECS PHOIBOS 150 
using linearly polarized synchrotron radiation from the beamlines 
UE112-PGM1 and UE112-lowE-PGM2 at BESSY II. Spin resolution was done with 
a Rice University Mott polarimeter operated at 26 kV [34]. The setup is 
sensitive to the two spin quantization axes in the surface plane of the 
sample, and Fig. 3j-l shows the spin component tangential to the energy 
surfaces which are shown in Fig. 3e-h. Scanning tunneling microscopy (STM) 
images were acquired with an Omicron VT-STM at room temperature.

\begin{center}
\vskip0.5cm
{\bf Sample characterization}
\vskip0.5cm
\end{center}

Prior to the detailed investigation of bare Ir(111), its cleanliness was 
verified by observation of a sharp {\it p}(1$\times$1) pattern in LEED (Fig. 1a) and 
by the absence of carbon-derived electronic states in valence-band 
photoemission. After covering the Ir with graphene, quality and 
completeness of the graphene layer were extensively tested. The samples 
have been characterized in situ with LEED (Fig. 1b) and in a separate 
setup with STM (Fig. S1a). The results of LEED and STM are in full 
agreement with previous reports [20]. 

The {\it in situ} characterization by ARPES 
showed typical features of the moir\'e pattern reported in the literature [27]: 
The Dirac cone is subject to lateral quantization by the moir\'e superlattice 
(Fig. S1b). Minigaps (dashes) appear where umklapp-induced replica bands 
(white arrows) intersect [27]. The characteristic trigonal symmetry of the 
Dirac cone as well as its two-dimensional replication within the surface 
Brillouin zone are well seen in constant binding energy surfaces (Fig. S1c).

\begin{center}
\vskip0.5cm
{\bf Topological nontrivial character of the surface state - theoretical support}
\vskip0.5cm
\end{center}

\begin{figure}[b]
\centering
\includegraphics[width=0.50\textwidth]{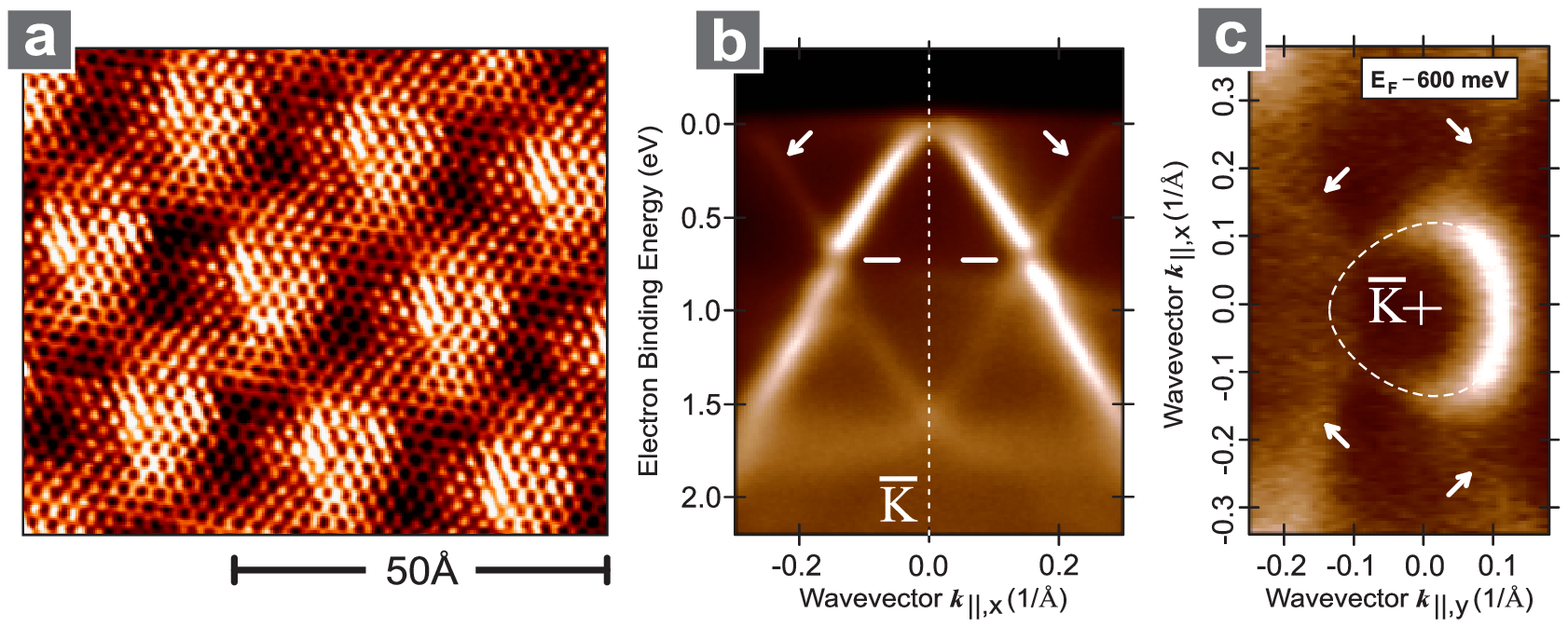}
%\caption{
\begin{flushleft}
{\bf Figure S1.}
Effects of the moir\'e-type superstructure of 
graphene/Ir(111) in ({\bf a}) STM and ({\bf b,c}) ARPES.
\end{flushleft}
%}
\end{figure}

We present here strong indications for a topological nontrivial character of the 
investigated surface state.  Fig. S2a shows the calculated band structure of Ir 
along [111] ($\Gamma$-$\Lambda$-L-direction) [35]. Unlike with noble metals, there 
is no band gap along [111] near E$_{F}$ because the red and the blue band overlap 
energetically in some part of the $\Lambda$-direction. As the present surface state is situated 
between the red and the blue band, it is strictly speaking a surface resonance. 
The red bulk band is of $\Lambda_{6}^{3}$ symmetry (subscript for double-group, superscript 
for single-group notation) and extends up to E$_{F}$ and above. The blue bulk band 
is a $\Lambda_{6}^{1}$ band with the noted energetical overlap with the $\Lambda_{6}^{3}$  
band in the range around 1 eV binding energy. These bands have opposite parities at 
the L point, L$_{6-}^{1}$ and L$_{6+}^{2'}$, respectively. From the band structure it appears 
thus that the surface state couples at L bulk states of odd parity and even 
parity which renders it odd, rather similar to the situation after band inversion 
in HgTe quantum wells creating a topological state as well [36].

\begin{figure*}[t]
\centering
\includegraphics[width=0.85\textwidth]{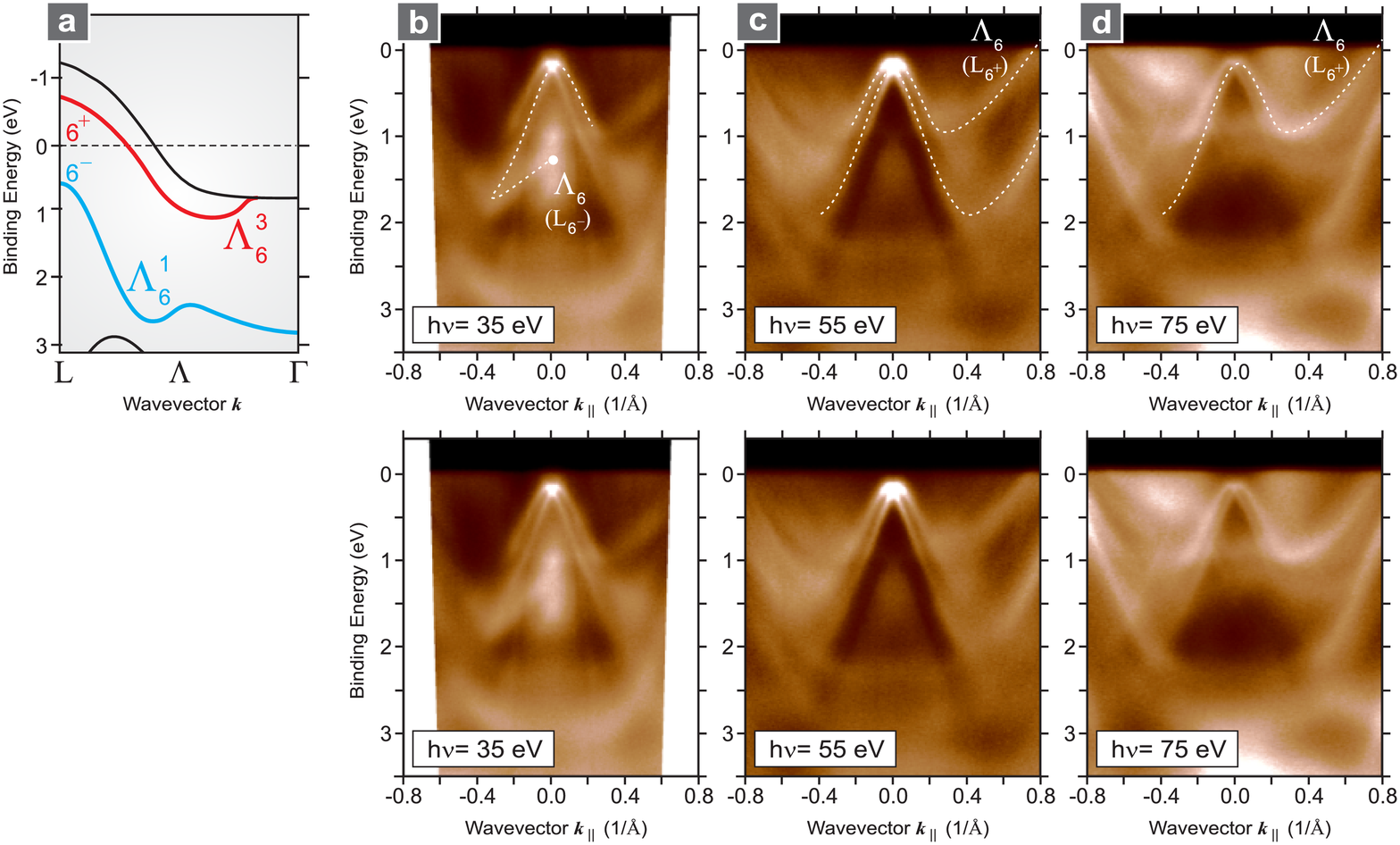}
%\caption{
\begin{flushleft}
{\bf Figure S2.}
({\bf a}) Calculated Ir bulk bands along [111] [35]. The surface state is situated 
in between the red and the blue band. ({\bf b-d}) Photon-energy and, hence, {\bf k}$_\perp$ 
dependence of the topological surface state and the adjacent bulk bands.
\end{flushleft}
%}
\end{figure*}

In Fig. S3a-b we show the criterion for topological protection of the surface 
state. Between two time-reversal invariant points of the surface Brillouin zone, 
i. e., here the two-dimensional $\overline{\Gamma}$ and $\overline{\rm M}$ points, the topological surface state 
must cross only an odd number of times an arbitrary line between the upper and 
lower bulk bands. This criterion replaces that for topological insulators with 
an absolute band gap at E$_{F}$ (Fig. S3a). Fig. S3b shows three different choices 
for such line proving that it cuts the surface state only an odd number of times.

\begin{center}
\vskip0.5cm
{\bf Topological nontrivial character of the surface state - experimental support}
\vskip0.5cm
\end{center}

Fig. 3a-c of the manuscript has shown that the surface state consists of entangled 
bands that cross at {\it k}$_\parallel$=0. This is very different from the finding 
of two parallel dispersions in Ref. 26. A possible reason could be a slight 
misalignment away from {\it k}$_\parallel$=0 in Ref. 26 because this would automatically 
lead to the disappearance of the crossing point. 

Fig. S2b-d [37] shows that the 
surface state does not change when the photon energy is varied from 35 eV over 
55 eV to 75 eV which changes the electron wave vector component perpendicular to the 
surface, {\bf k}$_\perp$. This proves its two-dimensional character.

\begin{figure}[b]
\centering
\includegraphics[width=0.47\textwidth]{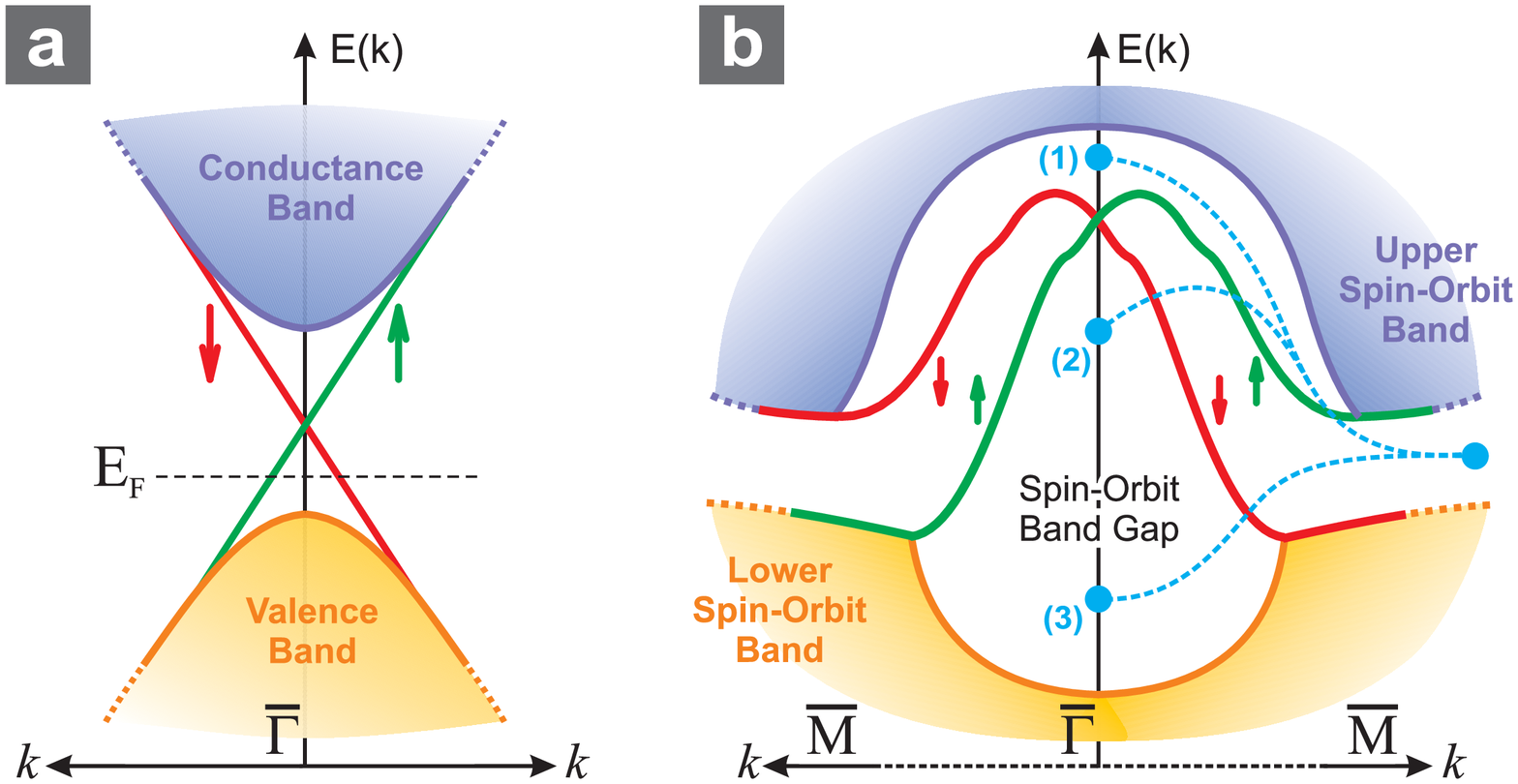}
%\caption{
\begin{flushleft}
{\bf Figure S3.}
Comparison of ({\bf a}) the topological insulator and ({\bf b}) the present case of 
a topological metal. (1), (2), and (3) denote three different choices for connecting 
time-reversal invariant points $\overline{\Gamma}$ and $\overline{\rm M}$ between the confining bulk bands.
\end{flushleft}
%}
\end{figure}

\begin{figure*}[t]
\centering
\includegraphics[width=0.95\textwidth]{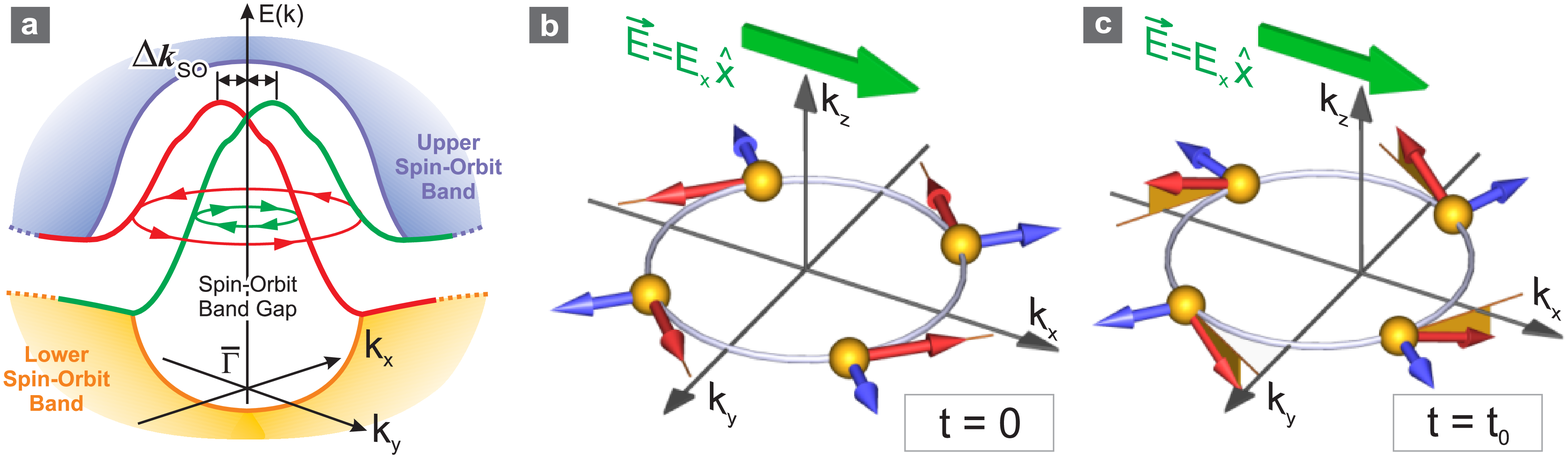}
%\caption{
\begin{flushleft}
{\bf Figure S4.}
({\bf a}) Sketch of the present Rashba-split topological surface state. 
({\bf b}) Constant-energy surface with directions of momentum (blue) and spin (red). 
({\bf c}) An external electric field leads to a spin tilting and spin 
current in the perpendicular in-plane direction, according to Ref. [4]. 
\end{flushleft}
%}
\end{figure*}

Other bands in Fig. S2b-d do, however, change strongly and are, therefore, derived 
from the three-dimensional bulk. In Fig. S2b we see that the inner cone of the surface 
state connects for negative {\it k}$_\parallel$ to a bulk band which reaches normal emission and 
thus the $\Gamma$-$\Lambda$-L line at 1.0-1.5 eV binding energy. From Fig. S2a we 
identify this bulk band as the lower $\Lambda_{6}^{1}$ band (blue) which connects 
to the L$_{6-}^{1}$ point of odd parity. Fig. S2c shows how at 55 eV the surface 
state connects for positive {\it k}$_\parallel$ to a bulk band which disperses 
upwards and crosses E$_{F}$. There are few bulk bands which cross E$_{F}$ [35]. 
The lower $\Lambda_{6}^{1}$ band does not cross E$_{F}$ but the upper $\Lambda_{6}^{3}$ 
band does so. So, even though the band in Fig. S2c is not sampled along the 
$\Lambda$ line, it is most likely that it is connected to this band. Fig. S2c 
also shows how the surface state connects to the lower bulk band (blue) at this 
higher photon energy and different value of {\bf k}$_\perp$ dispersing differently with 
{\bf k}$_\parallel$. Finally, Fig. S2d shows in the same data set that the surface 
state connects with its outer cone (for positive {\it k}$_\parallel$) to the upper 
bulk band (red) and with its inner cone (for negative {\it k}$_\parallel$) to the 
lower bulk band (blue).

Fig. S3 shows the relation of the present topological 
surface state on a metal to that on a topological insulator with absolute band gap. 
The surface state on a topological metal has at first been noted for Sb(111) [31]. 
For the topological insulator, an odd number of crossings of the Fermi energy 
between two time-reversal invariant points of the surface Brillouin zone is 
the necessary criterion. This means for the topological metal that any line 
connecting two time-reversal invariant points, here $\overline{\Gamma}$ 
and $\overline{\rm M}$, inside of the gap between the upper and lower bulk band 
must be crossed an odd number of times. This is shown for three examples (1) 
through (3) in Fig. S3b. 

\begin{center}
\vskip0.5cm
{\bf Principle of devices}
\vskip0.5cm
\end{center}

The device for creation of a spin current from a Rashba-split electronic 
structure is based on the intrinsic spin Hall effect. It is different from 
the historically older extrinsic spin Hall effect which is due to scattering 
from impurities [38]. The generation of spin currents does not require the 
band gap of a semiconductor. The generation and, not less difficult, the 
measurement of spin currents has made impressive progress in recent years. 
Currently, the spin Hall effect and inverse spin Hall effect have evolved 
into the most important processes for generation and measurement, 
respectively. The spin Hall effect has been proven at first for 
semiconductors [39] and for a two-dimensional electron gas [40-42]. 
However, the size of the effect in metals is much larger. It was shown 
that in Pt it is by 4 orders of magnitude larger than in semiconductors [43], 
and an FePt/Au device increased the effect further by another factor of 10 [44].

Fig. S4b shows how a spin current is created by a charge current in a system 
with Rashba-type spin-orbit interaction. The figure shows a sketch of a 
Rashba-split system adopted to the present situation from Ref. [4]. Due to the 
Rashba effect the electron momentum (blue) is coupled to the spin (red). For 
the case of circular constant energy surfaces as seen in the experimental data 
in Fig. 3e-h, it is tangential to these constant energy surfaces and perpendicular 
to the momentum. Fig. S4b shows for {\it t=0} the equilibrium situation when the external 
electric field is switched on. Fig. S4b shows further how the constant energy 
surface has shifted at {\it t}$_{0}$ along {\it k}$_{x}$ by  
$\Delta k_{x}=\frac{eE_{x}t_{0}}{\hbar^2}$. The movement leads in the 
perpendicular direction {\it k}$_{y}$ to a spin torque which tilts the spins up for 
{\it k}$_{y}>$0 and down for {\it k}$_{y}<$0. The contributions from the Fermi 
surface lead to a spin current in the y-direction. 

The effect of the graphene formation lifts the top of the surface state 
from 340 meV to 190 meV (Fig. 3a,b), so the surface state does not yet cut 
the Fermi energy. Alloying with a heavy element to the left of Ir in the 
perodic system such as Os for amounts of the order of 20\% should move the 
bulk band edges sufficiently up to make the surface state partially unoccupied. 

The active element of a potential device is the surface of Ir(111) which can 
be grown as a thin film. The role of the graphene is to protect this 
surface against ambient atmosphere. This is realistic as we have shown that 
the presence of the graphene leaves the Rashba splitting of the surface 
state unaffected and this does not change even after exposure to 
atmospheric pressure. 

\begin{center}
\vskip0.5cm
{\bf Stability of graphene and graphene on metals to atmosphere}
\vskip0.5cm
\end{center}

Graphene is stable at atmospheric pressure which is apparent from most 
experiments on exfoliated graphene flakes which are done under 
ambient conditions. If graphene would not be stable enough to protect the Ir 
surface against reactions with, e. g., oxygen,  this would ultimately change 
the bulk electronic structure of the Ir and also remove the surface state. 
In this sense, the stability and protection of graphene on transition metal 
surfaces is a prerequisite for the present work. While graphene/Ir has not 
yet been studied in this respect, other transition metals have. We have 
shown early on that exposure of graphene/Ni(111) does not lead to oxidation 
of the Ni [45,46]. The same system has been exposed to oxygen at 
6$\times$10$^{-6}$ mbar for 30 min [47], graphene/Fe/Ni(111) to similar 
conditions [48]. The system graphene/Ru(0001) has been exposed to oxygen 
partial pressure [49,50] and to atmosphere [50]. Surface electronic 
features of the transition metals have not been studied up to now.

%\onecolumngrid
\begin{center}
\vskip0.5cm
{\bf References}
\vskip0.5cm
\end{center}
\small

\renewcommand{\labelenumi}{[\theenumi]}

\begin{enumerate}
\setcounter{enumi}{33}

\item G. C. Burnett, T. J. Monroe, and F. B. Dunning, High-efficiency retarding-potential
Mott polarization analyzer, Rev. Sci. Instrum. 65, 1893 (1994).

\item
J. Noffke and L. Fritsche,  Band structure calculation and photoemission analysis of    
iridium, J. Phys. F.: Met. Phys. 12, 921 (1982).

\item
B. A. Bernevig, T. L. Hughes, and S.-C. Zhang, Quantum Spin Hall Effect and 
Topological Phase Transition in HgTe Quantum Wells, Science 314, 1757 (2006).

\item
J. S\'anchez-Barriga, A. Varykhalov et al. (unpublished)

\item 
J. E. Hirsch, Spin Hall effect, Phys. Rev. Lett. 83, 1834 (1999).

\item
Y. K. Kato, R. S. Myers, A. C. Gossard, and D. D. Awschalom, Observation of the spin     
hall effect in semiconductors, Science 306, 1910 (2004).

\item 
J. Wunderlich, B. Kaestner, J. Sinova, and T. Jungwirth, Experimental observation of the 
spin-Hall effect in a two-dimensional spin-orbit coupled semiconductor system, Phys. 
Rev. Lett. 94, 047204 (2005).

\item 
S. O. Valenzuela and M. Tinkham, Direct electronic measurement of the spin Hall effect, 
Nature 442, 176 (2006).

\item 
E. Saitoh, M. Ueda, H. Miyajima, and G. Tatara, Conversion of spin current into charge 
current at room temperature: Inverse spin-Hall effect, Appl. Phys. Lett. 88, 182509 
(2006).

\item 
T. Kimura, Y. Otani, T. Sato, S. Takahashi, and S. Maekawa, Room-temperature 
reversible spin Hall effect, Phys. Rev. Lett. 98, 156601 (2007).

\item 
T. Seki, Y. Hasegawa, S. Mitani, S. Takahashi, H. Imamura, S. Maekawa, J. Nitta, and 
K. Takanashi, Giant spin Hall effect in perpendicularly spin-polarized FePt/Au devices, 
Nature Mater. 7, 125 (2008).

\item 
V. A. Mozhayskiy, A. Y. Varykhalov, A. G. Starodubov, A. M. Shikin, S. I.  
Fedoseenko, and V. K. Adamchuk, Formation of mono-atomic carbon layers on Ni(111) 
by means of organic-gas cracking and by thermal decomposition of fullerenes in thin 
film, Phys. Low-Dim. Struct., Issue 1-2, 105 (2003). 

\item 
V. A. Mozhayskiy, A. Y. Varykhalov, A. G. Starodubov, A. M. Shikin, S. I. 
Fedoseenko, and V. K. Adamchuk, Two alternative ways for formation of mono-atomic    
carbon layer on Ni(111): Organic-gas cracking and thermal decomposition of fullerenes 
in thin film, Fullerenes, Nanotubes and Carbon Nanostructures 12, 385 (2004). 

\item 
Y. S. Dedkov, M. Fonin, and C. Laubschat, A possible source of spin-polarized 
electrons: The inert graphene/Ni(111) system, Appl. Phys. Lett. 92, 052506 (2008).

\item 
Y. S. Dedkov, M. Fonin, U. R\"udiger, and C. Laubschat, Graphene-protected iron layer on 
Ni(111), Appl. Phys. Lett., Appl. Phys. Lett. 93, 022509 (2008).

\item 
H. Zhang, Q. Fu, Y. Cui, D. Tan, and X. Bao, Growth Mechanism of Graphene on 
Ru(0001) and O2 Adsorption on the Graphene/Ru(0001) Surface, J. Phys. Chem. C 113, 
8296 (2009).

\item 
B. Borca, F. Calleja, J. J. Hinarejos, A. L. V\'azquez de Parga, and R. Miranda, Reactivity 
of periodically rippled graphene grown on Ru(0001), J. Phys.: Condens. Matter 21, 
134002 (2009).

\end{enumerate}


\begin{thebibliography}{1}


\bibitem{Rashba}
E. I. Rashba, Fiz. Tver. Tela (Leningrad) 2, 1224 (1960), 
[Properties of semiconductors with an extremum loop. 1. Cyclotron and combinational 
resonance in a magnetic field perpendicular to the plane of the loop,
Sov. Phys. Solid State 2, 1109 (1960)]; 
Y. A. Bychkov and E. I. Rashba, 
Oscillatory effects and the magnetic susceptibility of carriers in inversion layers,
J. Phys. C 17, 6039 (1984).

\bibitem{Winkler}
R. Winkler, Spin-Orbit Coupling Effects in Two-Dimensional Electron and Hole Systems,
 Springer, Berlin (2003).

\bibitem{MurakamiScience03}
S. Murakami, N. Nagaosa, and S.-C. Zhang, 
Dissipationless quantum spin current at room temperature,
Science 301, 1348 (2003).

\bibitem{SinovaPRL04}
J. Sinova, D. Culcer, Q. Niu, N. A. Sinitsyn, T. Jungwirth, and A. H. MacDonald,
Universal Intrinsic Spin Hall Effect, 
Phys. Rev. Lett. 92, 126603 (2004). 

 \bibitem{online}
See supporting online material.

%\bibitem{Nitta97}
%J. Nitta, T. Akazaki, H. Takayanagi, and T. Enkoki, 
%Gate Control of Spin-Orbit Interaction in an Inverted 
%In0.53Ga0.47As/In0.52Al0.48As Heterostructure,
%Phys. Rev. Lett. 78, 1335 (1997). 
% Kommentar zu alpha_R siehe unter Dil

%\bibitem{MatsuyamaPRB00}
%T. Matsuyama, R. K\"ursten, C. Meissner, and U. Merkt,
%Rashba spin splitting in inversion layers on p-type bulk InAs
%Phys. Rev. B 61, 15588 (2000). 

\bibitem{Wu10}
M. W. Wu, J. H. Jiang, and M. Q. Weng, 
Spin dynamics in semiconductors,
Phys. Rep. 493, 61 (2010).

\bibitem{TKimura}
T. Kimura, Y. Otani, T. Sato, S. Takahashi, and S. Maekawa, Room-temperature 
  reversible spin Hall effect, Phys. Rev. Lett. 98, 156601 (2007).

\bibitem{Seki}
T. Seki, Y. Hasegawa, S. Mitani, S. Takahashi, H. Imamura, S. Maekawa, J. Nitta, and 
  K. Takanashi, Giant spin Hall effect in perpendicularly spin-polarized FePt/Au devices, 
  Nature Mater. 7, 125 (2008).

\bibitem{Tamm}
I. Tamm, 
\"Uber eine m\"ogliche Art der Elektronenbindung an Kristalloberfl\"achen, 
Physik. Z. Sowjetunion 1, 733 (1932).
% I., Ig., I. E., I. J., I. Y.,

\bibitem{ReviewPlummer} 
E. W. Plummer and W. Eberhardt, 
Angle-resolved photoemission as a tool for the study of surfaces, 
Adv. Chem. Phys. 49, 533 (1982).

\bibitem{LaShell} 
S. LaShell, B. A. McDougall, and E. Jensen,
Spin splitting of an Au(111) surface state band observed
with angle resolved photoelectron spectroscopy,
Phys. Rev. Lett. 77, 3419 (1996); 
G. Nicolay, F. Reinert, S. H\"ufner, P. Blaha, 
Spin-orbit splitting of the L-gap surface state on Au(111) and Ag(111),
Phys. Rev. B 65, 033407 (2001);
M. Hoesch, M. Muntwiler, V. N. Petrov, M. Hengsberger, L. Patthey, M. Shi, M. Falub, 
T. Greber, and J. Osterwalder, 
Spin structure of the Shockley surface state on Au(111),
Phys. Rev. B  69, 241401(R) (2004).

\bibitem{Hedegard} 
L. Petersen and P. Hedeg\aa rd, 
A simple tight-binding model of spin-orbit splitting
of sp-derived surface states,
Surf. Science 459, 49 (2000).
 
\bibitem{bismuth}
T. Hirahara, K. Miyamoto, I. Matsuda, T. Kadono, A. Kimura, T. Nagao,
 G. Bihlmayer, E. V. Chulkov,  S. Qiao, 
K. Shimada, H. Namatame, M. Taniguchi,  and S. Hasegawa,
Direct observation of spin splitting in bismuth surface states,
Phys. Rev. B 76, 153305 (2007).

\bibitem{Hochstrasser-PRL-2002} 
M. Hochstrasser, J.G. Tobin, E. Rotenberg, and S. D. Kevan, 
Spin-resolved photoemission of surface states of W(110)-$(1\times1)$H,
Phys. Rev. Lett. 89, 216802 (2002).

\bibitem{RotenbergLiW110} 
E. Rotenberg, J. W. Chung, and S. D. Kevan,
Spin-orbit coupling induced surface band splitting in Li/W(110) and 
Li/Mo(110),
 Phys. Rev. Lett. 82, 4066 (1999).

\bibitem{Shikin-PRL-2008}
A. M. Shikin, A. Varykhalov, G. V. Prudnikova, D. Usachov, V. K. Adamchuk, 
Y. Yamada, J. D. Riley, and O. Rader,
Origin of spin-orbit splitting for monolayers of Au and Ag on W(110) and Mo(110),
Phys. Rev. Lett. 100, 057601 (2008).

\bibitem{Varykhalov-QWS-PRL-2008}
A. Varykhalov, J. S\'anchez-Barriga, A. M. Shikin, W. Gudat, W. Eberhardt, 
and O. Rader,
Quantum cavity for spin due to spin-orbit interaction at a metal boundary,
Phys. Rev. Lett.   101, 256601 (2008).

\bibitem{Dil-PRL-2008}
J. H. Dil, F. Meier, J. Lobo-Checa, L. Patthey, G. Bihlmayer, and J. Osterwalder,
Rashba-type spin-orbit splitting of quantum well states in ultrathin Pb films,
Phys. Rev. Lett.  101, 266802 (2008).
% Pb/Si(111), alpha_R=0.04 +/-0.005 eV*Angstr in expt. and 0.071 eV*Angstr. in theory
% 0.04eV*Angstr = 0.04E-10 eVm = 4E-12 eVm
% Au111: 0.33 eV*Angstr = 3.3E-11 eVm
% Nitta 0.07 eV*Angstr = 7E-12 eVm (aber delta-k ist �hnlicher)

\bibitem{Varykhalov-Graphene-PRL-2008}
A. Varykhalov, J. S\'anchez-Barriga, A. M. Shikin, C. Biswas, E. Vescovo, 
A. Rybkin, D. Marchenko, and O. Rader,
Electronic and Magnetic Properties of Quasifreestanding Graphene on Ni,
Phys. Rev. Lett. 101, 157601 (2008).

\bibitem{Ast} 
Ch. S. Ast, J. Henk, A. Ernst, L. Moreschini, M. C. Falub, D. Pacil\'e, 
K. Kern and M. Grioni,
Giant spin splitting through surface alloying,
Phys. Rev. Lett. 98, 186807 (2007).
% room temp. delta-K=0.12invA and at 77K it is 0.16invA

\bibitem{NDiaye-NJP-2008}
A. T. N'Diaye, J. Coraux, T. N. Plasa, C. Busse, T. Michely,
Structure of epitaxial graphene on Ir(111),
New J. Phys. 10, 043033 (2008).

\bibitem{VeenSCLS}
J. F. van der Veen, F. J. Himpsel, and D. E. Eastman,
Structure-dependent 4f-core-level binding-energies for surface atoms on Ir(111),
Ir(100)-$(5\times1)$, and metastable Ir(100)-$(1\times1)$, 
Phys. Rev. Lett. 44, 189 (1980).
% -0.50 eV shift

\bibitem{BianchiNJP09}
M. Bianchi, D. Cassese, A. Cavallin, R. Comin, F. Orlando,
L. Postregna, E. Golfetto, S. Lizzit and A. Baraldi,
Surface core level shifts of clean and oxygen covered Ir(111),
New J. Phys. 11, 063002 (2009).
% 550meV shift
%"Ir 4f7/2 core level spectra are shown to be very sensitive to the 
%local atomic environment" "Oxygen adsorption at low coverage leads to the
%appearance of a new component Ir1 which grows between the surface 
%and the bulk peaks, while the intensity of the original Ir0 surface peak decreases.
 
\bibitem{Lacovig-PRL-2009}
P. Lacovig, M. Pozzo, D. Alf\'e, P. Vilmercati, A. Baraldi, S. Lizzit,
Growth of Dome-Shaped Carbon Nanoislands on Ir(111): 
The Intermediate between Carbidic Clusters and Quasi-Free-Standing Graphene, 
Phys. Rev. Lett. 103, 166101 (2009).

\bibitem{VeenSSPRB80}
J. F. van der Veen, F. J. Himpsel, and D. E. Eastman,
Experimental energy dispersions for valence and conduction bands of iridium,
Phys. Rev. B 22, 4226 (1980).
  
\bibitem{PletikosicJPCM10} 
I. Pletikosi\'c, M. Kralj, D. Sokcevi\'c, R. Brako, P. Lazi\'c, and P. Pervan, 
Photoemission and density functional theory study of Ir(111); energy band gap mapping,
J. Phys.: Condens. Matter 22, 135006 (2010).
% sie zeigen zwei parallele SS
%  Sokcevi\'c mit umgekehrtem Hut auf S und erstem c.

\bibitem{Pletikosic-PRL-2008}
I. Pletikosi\'c, M. Kralj, P. Pervan, R. Brako, J. Coraux, A. T. N�Diaye, C. Busse, 
T. Michely,
Dirac Cones and Minigaps for Graphene on Ir(111),
Phys. Rev. Lett. 102, 056808 (2009).

\bibitem{Giovannetti}
G. Giovannetti, P. A. Khomyakov, G. Brocks, V. M. Karpan, J. van den Brink, 
and P. J. Kelly, 
Doping graphene with metal contacts,
Phys. Rev. Lett. 101, 026803 (2008).

\bibitem{topological}
C. L. Kane and E. J. Mele, 
Z$_2$ topological order and the quantum spin Hall effect, 
Phys. Rev. Lett. 95, 226801 (2005); 
B. A. Bernevig and S. C. Zhang, 
Quantum spin Hall effect, 
Phys. Rev. Lett. 96 106802 (2006);
M. K\"onig {\it et al.}, 
Quantum spin hall insulator state in HgTe quantum wells,
Science 318, 766 (2007); 
J. E. Moore and L. Balents, 
Topological invariants of time-reversal-invariant band structures, 
Phys. Rev. B 75, 121306 (2007); 
R. Roy, 
Topological phases and the quantum spin Hall effect in three dimensions, 
Phys. Rev. B 79, 195322 (2009); 
S. Murakami, 
Phase transition between the quantum spin Hall and insulator phases in 3D: 
emergence of a topological gapless phase,
New J. Phys. 9,  356 (2007).
 
\bibitem{XiaNatPhys}
Y. Xia, D. Qian, D. Hsieh, L. Wray, A. Pal, H. Lin, A. Bansil, D. Grauer, 
Y. S. Hor, R. J. Cava,
and M. Z. Hasan, 
Observation of a large-gap topological-insulator class with a single Dirac cone
on the surface, 
Nature Phys. 5, 398 (2009).

\bibitem{HsiehScience09}
D. Hsieh {\it et al.}, Observation of Unconventional Quantum Spin Textures 
in Topologically Ordered Materials, 
% D. Hsieh, Y. Xia, L. Wray, D. Qian, A. Pal, J. H. Dil, F. Meier, 
% J. Osterwalder, G. Bihlmayer, C. L. Kane, Y. S. Hor, R. J. Cava, 
% M. Z. Hasan
Science 323, 919 (2009).

\bibitem{PLDS}
V. A. Mozhayskiy, A. Y. Varykhalov, A. G. Starodubov, A. M. Shikin, S. I. Fedoseenko, 
and V. K. Adamchuk, 
Formation of mono-atomic carbon layers on Ni(111) by means of organic-gas 
cracking and by thermal decomposition of fullerenes in thin film,
Phys. Low-Dim. Struct., Issue 1-2, 105 (2003). 
See Supplementary Material for more recent references.


\bibitem{Supplement}
Supplementary information for this paper is enclosed.

 
\end{thebibliography}
\end{document}